\documentstyle[pra,preprint,aps,epsf,amsfonts]{revtex}
\tighten
\begin{document}
\draft
\title{Operational Time of Arrival in Quantum Phase Space}
\author{Piotr Kocha\'nski}
\address{Centrum Fizyki Teoretycznej PAN,
Al. Lotnik\'ow 32/46, 02-668 Warszawa, Poland}
\author{Krzysztof W\'odkiewicz\cite{unm}}
\address{Instytut Fizyki Teoretycznej, Uniwersytet Warszawski,
Ho\.{z}a~69, 00--681~Warszawa, Poland}
\date{version {\tt time.tex}; \today}
\maketitle
\begin{abstract}
An operational time of arrival is introduced using a realistic 
position and momentum measurement scheme.
The phase space measurement involves the dynamics of a quantum particle
probed by a measuring device.
For such a  measurement an operational
positive operator valued measure  in phase space is introduced and 
investigated.
In such an operational formalism a quantum mechanical
time operator is constructed and analyzed. A phase space time and
energy uncertainty relation is derived.

\end{abstract}

\section{Introduction}
The problem of the proper definition  of the quantum mechanical time
observable, and the physical interpretation of the associated
uncertainty relation between time and energy, is a subject of a long 
lasting debate. This is due to the fact, that both in classical 
and quantum nonrelativistic  mechanics, the
time parameter is not an independent dynamical variable, and its definition
requires the use of a nonlinear function of the canonical variables
describing the system. In classical physics the form of this nonlinear
function does not bring any conceptual difficulties, for instance, for a free
particle of a unit mass and in one dimension, with an initial momentum
$p_{0}=p(t_{0})$ and an initial position $q_{0}=q(t_{0})$, the time of arrival
from $q_{0}$ to a fixed position $q$  is given by
\begin{equation}
\label{classtime}
t=t_0+{q-q_{0} \over p_{0}}\,.
\end{equation}
In the  general case of an arbitrary motion it is necessary to invert
the equations of motion for a particle $q=q(t,q_{0},p_{0})$ in order to find
the time parameter $t=t(q,q_{0},p_{0})$. This function can be in general
multivalued and the choice of the form $t$ is selected  according to
the physical meaning of the given solution.

In quantum mechanics the problem of the definition of the time
observable  associated with such a nonlinear transformation  becomes much
more complicated since canonical observables are noncommuting operators.
This makes, for instance, the definition like (\ref{classtime}) ambiguous due
to the ordering problem of $\hat q$ and $\frac{1}{\hat p}$ operators.

It was Pauli, who first originated  the question whether 
one can define a time operator as a canonical variable conjugated to the
energy of the system. In his book on quantum mechanics 
\cite{pauli33}, he
pointed out, that although formally we can write
$[\hat{H},\hat{t}]=i\hbar$, this formula is unsatisfactory
since the operator $\hat{t}$ cannot be Hermitian due to the
semi -- bounded character of the Hamiltonian spectrum.
As a result, the uncertainty relation in the form
$\Delta E \Delta t\geq\hbar$ has  a disputable
interpretation.

One can derive, however, the above uncertainty relation using a Fourier
decomposition of a nonstationary  state in the form
 $\psi(x,t)$$=$$\int_{0}^{\infty} dE\psi(x,E)\exp{(-iEt/\hbar)}$.
In such an approach, the uncertainty relation  is a simple consequence of
Fourier analysis, similar to the one encountered  in classical Fourier optics
for  time-dependent fields and their spectra. A similar argument can be
applied to derive the  position and the momentum uncertainty relation of a
wave packet in wave mechanics. However, it is well
known that this last uncertainty relation has a much deeper meaning being the
consequence of the fundamental quantum incomplementarity of two 
canonical variables -- momentum and position.
Such an approach to quantum nonstationary states was presented first by
Fock and Krylov \cite{vf+nk47}, and in the basically
same spirit by Mandelstamm and Tamm \cite{lm+it45}. In these works 
the relation between the energy of the system and
its ``inner'' time \cite{ya+db6164}, which  usually is  the lifetime of 
the nonstationary quantum state, has been derived and discussed

Another approach was proposed by Landau and Peierls \cite{ll+rp31}
and  by Fock and Krylov  \cite{vf+nk47} who  discussed the problem of the
time-energy uncertainty relation analyzing a
typical scattering experiment of a particle  on a test particle.
The essential conclusion of these contributions was that the uncertainty
relation is connected with the inability of measuring
precisely the energy of a given system in an arbitrary short period of time.

Neither of the above proposals lead to the definition of the time
operator, what's more, it has been shown by Aharonov and Bohm
\cite{ya+db6164} that the interpretation given by
Landau and Peierls is erroneous i.e., it is possible
to measure the energy in a time period which violates
the postulated uncertainty relation. In the same article
Aharonov and Bohm  proposed a different approach to the problem
of the time observable. They argued that in order to measure the time of
arrival to a certain fixed point one needs to have a ``quantum
clock'' i.e., another quantum system, which  reads the experimental data
collected during a clock measurement.
This assumption led to the first explicit definition of the time observable
in the form proposed by Aharonov and Bohm:
\begin{equation}
\label{taharonov}
\hat{T}_c\equiv\frac{1}{2}\left(\hat{q}\frac{1}{\hat{p}}+
\frac{1}{\hat{p}}\hat{q}\right),
\end{equation}
where $\hat{q},\hat{p}$ are the operators of the clock particle.
This definition formally leads to the commutation relation
$[\hat{H_c},\hat{T}_c]=i\hbar$, where ${\hat H}_c$ is the Hamiltonian
of a clock particle. However such a time operator  has an important
disadvantage, in an a priori imposed and physically unjustified
choice of a symmetric ordering of  momentum and  position operators.
On the other hand, such an operator has a number of
interesting properties \cite{ng+cr+rst96,jgm+jpp+crl98}
that indicate possible  physical applications.

This fact can be understood in two ways.. First,
it has been shown in Ref. \cite{pb+mg+pjl94,gianni97}, that with the Aharonov -- Bohm time
operator (\ref{taharonov}) one can associate  a 
positive operator valued measure (POVM), or equivalently,
a realistic measurement scheme, whose outcomes are described
by such an observable.  

Another interpretation of the Aharonov -- Bohm formula (\ref{taharonov})
might be given in the framework of the approach proposed
by Kijowski \cite{kij74}. It was shown there, that by a construction of
a probability distribution for a particle to pass at
a certain moment a two dimensional reference plane, one can
define a time operator. 
The explicit formula for such a time operator is 
 like (\ref{taharonov}):
\begin{equation}
\label{kijowski}
\hat{T}_K\equiv {\rm sgn}(\hat{p})\frac{1}{2}\left(\hat{q}\frac{1}{\hat{p}}+
\frac{1}{\hat{p}}\hat{q}\right),
\end{equation}
where the multiplicative sign of the momentum operator
assures the Hermitian character of the time observable \cite{kij98}.
Physically, the sign function  means that one distinguishes between
particles moving to the left and to the right with respect to chosen
two dimensional reference plane.
For an initial beam of particles
prepared in such a way, that it contains only particles moving
either to the left or to the right, the  time operator
(\ref{kijowski}) reduces to the original Aharonov and Bohm
formula (\ref{taharonov}). One should  however note that  this time
operator is built from the operators describing the measured
particle, and not from some clock particle operators like in the Aharonov --
Bohm approach. 

The time operator (\ref{kijowski})  is conjugated to the 
variable ${\rm sgn}(\hat{p})(\hat{p}^2 /2)$, and as a consequence
leads to a corresponding uncertainty relation \cite{kij74,kij98}.

It has been shown in \cite{vd+jgm97,vd97+98} that one can use the above
 definition as a starting point to look for  a time observable. 
The definitions of the time observable given by Kijowski \cite{kij74,kij98} 
and by Delgado and Muga \cite{vd+jgm97,vd97+98} are then equivalent.

It shall also be stressed, that contrary to the claims of some authors
\cite{ng+cr+rst96,vd+jgm97}, the time operator (\ref{kijowski}) is defined 
in the framework of standard quantum mechanics, although in
a rather unusual representation of the wave function, which
is, however, as good as any other representation (see comment \cite{kij98}).

It is also worth noticing that a symmetrized operator like 
(\ref{taharonov}) is naturally connected with a probability distribution
of arrival times derived from an expectation value of the properly defined
positive current operator \cite{vd+jgm97,vd97+98}.

After a number of early, classical papers, which stated the
problem and revealed the main difficulties in dealing
with the   question  of the  time observable in quantum mechanics,
there was a large number of further works.
Although our brief review of the published literature is far from being
complete, we would like briefly to describe several approaches in order to 
place
our own work in some historical and logical context. An excellent 
study, of both older and newest researches connected with the problem
of time observable in quantum mechanics, can be found in \cite{jgm+rs+jpp98}.

Roughly speaking, the majority of the research efforts  in this field
can be divided into the following three approaches.

In the first approach the main effort was concentrated on the proper
definition and understanding of various time and  energy uncertainty
relations. In such relations,  the meaning of both time and energy, varied
depending on the author and the presented method.  The most complete, 
known to us, review of this problem can be found in Ref. \cite{mb+pam78}.
Newest investigations on this issue have been presented in
Ref. \cite{ya+etal97}. 

Another way of dealing with  the time observable was
to define a time operator as a symmetrized function
of the position and  momentum operators 
\cite{ng+cr+rst96,jgm+jpp+crl98}. This approach was discussed above 
and is justified by the work of Kijowski.

The third method, the closest to our proposal, is based on the investigation
of a particle, whose time of arrival to a given point is detected by a
certain reference ``quantum clock''. This quantum clock is represented by
another quantum system, which has some  properties enabling us to read the
time of arrival of the measured particle. 
Obviously, one can design  many different models of such
quantum clocks. Depending how one chooses the system representing the
quantum clock, and how one models the interaction
between the clock and the investigated particle different nonequivalent
schemes of time measurement can be introduced.
Examples of various ``quantum clocks'' can be 
found in \cite{pb+mg+pjl94,ya+etal97,ap80,ya+jlf75}. 

Another interesting discussion of the realistic, irreversible,
detector model has been presented by Halliwell \cite{halliwell98}. The
decoherent histories approach to the arrival time problem has been
given by the same author in Ref. \cite{jjh+ez97}. In this paper a probability
distribution to find a particle in a certain position in a given
fixed time {\em interval} has been introduced, 

using similar idea presented earlier by Yamada and Takagi \cite{y+t91}.

This has to be
contrasted with the standard quantum approach, when such a probability is 
introduced only for a  given moment of time.
A similar approach, in which the detector is not specified as another
quantum system, but rather is treated phenomenologically without
explicit treatment of the interaction of the measured system
and the experimental device, is given in Ref. \cite{jgm+jpp+crl}.

One can also ask, if it is possible to choose among the different
quantum clocks those which are the most precise, in a sense that
they saturate the time -- energy uncertainty relation. This
question has been addressed in Ref. \cite{vb+rd+sm98}.  

A very extensive and general analysis of various aspects of
the proper treatment of the time observable has been given
in the papers by Allcock \cite{alloc69} and by Mielnik \cite{mielnik94}.

It is the purpose of this paper to define an arrival time operator based on 
a realistic momentum and position measurement. Our work is mainly motivated
by a recent successful experimental approach, in which operational
measurements of  the quantum phase of optical fields have been performed
\cite{nfm}. Such  operational measurements have been related to simultaneous
{\em joint} measurements of position and momentum operators in a space {\em
extended} by the measuring device.

In this paper  we define an operational time operator, which is
connected with a very simple model of a joint momentum and position
measurement. Such a measurement may be implemented
\cite{kw84,bge+kw95,blm,peres}, if  additional degrees of freedom responsible
for the measuring devices are involved. 
For such a measurement we introduce the operational
time observable and a corresponding POVM. We show that for a special
case of the quantum clock state, the arrival time operator
involves an antinormal ordering of the creation and annihilation operators
forming the canonical variables.
The operational phase space measurement leads to a time and energy
phase space. For such a phase space  and for an  operational time of arrival 
a time and energy uncertainty relation is derived.

This paper is organized in the following way. A short review
of the general  operational approach will be discussed in Section II.
In Section III we present  a quantum model for a joint
measurement of position and momentum in a space extended by the measuring
device.
This  allows  to construct a
corresponding POVM relevant to the definition of the operational time
observable. From this POVM  a {\em positively} defined
operational probability distribution
function called {\em propensity} is derived.
This propensity leads  to a mean time of arrival. 
In Section IV we introduce an explicit form of the operational
time of arrival. We derive the time operator and
show that the specific ordering of the position and momentum operators, 
forming this observable, depends on the properties of the measuring device.
Using the time operator, constructed in such a way, we
derive in Section V an operational time and energy uncertainty principle.
In Section VI we introduce a time and energy phase space. 
We show that an arrival time distribution can be interpreted in terms
of a positive and negative flux flow.
Finally 
Section VII contains some concluding remarks.

Through the rest of the paper we shall work in one dimension and use 
the units in which $\hbar=1$, $m=1$.

\section{Operational Quantum Measurement}

\subsection{Operational POVM}

In order to describe realistic experiments involving measuring devices  it is
sometimes necessary to go beyond the idealized measurement scheme proposed by
von Neumann \cite{vneumann32}.
In this approach it is postulated, that statistical outcomes of a measurement
of a certain observable $\hat{A}$ with eigenvalues and eigenvectors
$\hat{A}|a\rangle=a|a\rangle$ are described by the spectral measure:
\begin{equation}
\label{intrinsic_probability}
p_{\psi}(a)=|\langle a|\psi\rangle|^{2}\,,
\end{equation}
where $|\psi\rangle\in {\cal H}$ is the state vector of the measured system.
It is known that the spectral measure contains all the relevant statistical
informations about the investigated system but it makes no reference to the
apparatus employed in the actual measurement.
Due to this property $\hat{A}$ will be called an {\em intrinsic quantum
observable}.
In order to provide a  more realistic approach to the quantum measurement
we need to analyze carefully the dynamics of the combined system involving
{\em both} the measured system  and the
measuring apparatus. We shall call for short the measuring apparatus and all
its associated devices a {\em filter} device. 
Such an approach has been presented by many authors
\cite {kw84,bge+kw95,blm,peres} and we 
shall use here the formulation and notation from \cite{kw84,bge+kw95}.

Let the state of the measured system be described by the density
operator $\hat{\varrho}$  in a Hilbert space
${\cal H}$ and the state of filter device by
$\hat{\varrho}_{{\cal F}}$ in a Hilbert space ${\cal H}_{{\cal F}}$.
The evolution of the combined system, including the interaction between the
filter and the investigated system necessary for the measurement to happen,
is given by a unitary operator $\hat{U}(t,t_{0})$ acting in ${\cal
H}\otimes{\cal H}_{{\cal F}}$:
\begin{equation}
\hat{\varrho}(t_{0})\otimes \hat{\varrho}_{{\cal F}}(t_{0})\longrightarrow
\hat{U}(t,t_{0})\hat{\varrho}(t_0)\otimes
\hat{\varrho}_{{\cal F}}(t_0)\hat{U}^{\dagger}(t,t_{0})\,,
\,\,\,\, (t>t_{0}).
\end{equation}
In order to probe the system state with  such an interaction, the
statistical readouts of the filter device are collected.  The corresponding
probability distribution corresponds to a propensity of the measured state to
take on certain states of the filter. Due to the interaction the propensity
of the probed state to  be in one of the filter states
$|a\rangle $ is
\begin{equation}
\label{propensity}
\Pr(a)={\rm Tr}_{{\rm sys}}\{ \hat{\varrho} 
{\rm Tr}_{{\cal F}}[\varrho_{{\cal F}}
\hat{U}^{\dagger}|a\rangle\langle a|\hat{U}]\}
\equiv{\rm Tr}_{\rm sys}\{\hat{\rho}\hat{{\cal F}}(a)\},
\end{equation}
where $\hat{{\cal F}}(a)$ is a filter dependent POVM  satisfying the
normalization condition
\begin{equation}
 \int 
da\,\hat{{\cal F}}(a)=\hat{1}.
\end{equation}

\subsection{Operational operators}

In view of the linear relation between the propensity and the POVM, the
operational statistical moments of the measured quantity
\begin{equation}
\label{moments}
\overline{a^{n}}=\int da\,a^{n}\Pr(a)=\langle \hat{A}^{(n)}_{{\cal F}}\rangle
\end{equation}
define uniquely a set of {\em operational operators}
\begin{equation}
\label{op_operator}
\hat{A}^{(n)}_{{\cal F}}=\int da\,a^{n}\hat{{\cal F}}(a).
\end{equation}
A real number $a$ is a classical quantity which enables
us to read out the result of the quantum mechanical measurement.
Since a real measurement operates in a classical world
the outcomes of such an experiment are known only if they are
macroscopically recorded \cite{vkampen}. This is  why the result of such
an experiment will be given by the probability
distribution of a certain classical variable. In the presented
approach this distribution is just the quantum propensity $\Pr(a)$.

The operators (\ref{op_operator}) are called  operational operators,
because they
represent quantities measured in a real experiment involving a
dynamical coupling of the measured system with the filter. We see that in a
realistic measurement, with  a filter, the spectral
decomposition of an observable  $\hat{A}$ is effectively replaced by a
POVM $da\,\hat{{\cal F}}(a)$ \cite{blm}.
As a rule, the algebraic properties of the set $\hat{A}^{(n)}_{{\cal F}}$
differ significantly from those of the powers of $\hat{A}$.
For instance,
$(\hat{A}^{(1)}_{{\cal F}})^{2}\neq\hat{A}^{(2)}_{{\cal F}}$. This property
will have important consequences in the discussion of the uncertainty
introduced by the measurement. As an example  the operational
spread of a measured quantity, described by the 
of operational operators,  is given by
\begin{equation}
\delta A \equiv \langle \hat{A}^{(2)}_{{\cal F}} \rangle -
\langle \hat{A}^{(1)}_{{\cal F}}\rangle^{2} = \
\overline{a^2}-\overline{a}^2\,.
\end{equation}
This relation will play an important role in the formulation of the 
operational
uncertainty relation.

It is seen that the propensity and the operational operators
are natural extension of the spectral probability distribution and the
intrinsic operators. The difference is that the operational observables
carry information  about the system under investigation and the selected
measuring device. There always exists a physical mechanism (a realizable
experimental procedure) generating any desired POVM. This is guaranteed by
the Naimark extension, and the  reduction of the projection measure into the
Hilbert space of the measured system \cite{blm}. According to this theorem 
one can always extend the Hilbert space with the defined POVM to a space 
with a projective measure.

\section{Joint Momentum and Position Measurement}

\subsection{Joint measurements in phase space}

In this section we present a simple model of a joint
position and momentum measurement proposed originally by
von Neumann \cite{vneumann32} and  elaborated later
in Ref. \cite{ea+jlk65,ss92,pk+kw97}.

The interaction Hamiltonian between the
system (a particle) and  two filter particles (labeled by $1$ and $2$) is
given by
\begin{equation}
H_{I}=\delta(t)(\hat{q}\hat{p}_{1}-\hat{p}\hat{q}_{2}).
\end{equation}
After the pulsed interaction with the system, the measured
readouts $\hat{q}_{{\cal M}}$ and
$\hat{p}_{{\cal M}}$ of the filter variables $\hat{q}_{1}$ and 
$\hat{p}_{2}$ are:
\begin{eqnarray}
\label{naimark}
\hat{q}_{1} \longrightarrow \hat{q}_{{\cal M}}&=&\hat{U}^{\dagger}{\hat
q}_{1} \hat{U} =\hat q
+{\hat q}_{1}- {\hat q}_{2}/2\ ,  \nonumber\\
\hat{p}_{2} \longrightarrow \hat{p}_{{\cal M}}&=&\hat{U}^{\dagger}{\hat
p}_{2} \hat{U} =\hat p
+{\hat p}_{2}- {\hat p}_{1}/2\ .
\end{eqnarray}
In the combined space of the two filter particles, we introduce  filter
operators ${\hat q}_{{\cal F}}={\hat q}_{2}/2-{\hat q}_{1}$ and
$ {\hat p}_{{\cal F}}={\hat p}_{2}-{\hat p}_{1}/2$. In the product space of
the system and the filter the two measured observables (\ref{naimark}) are:
\begin{equation}
\label{epr}
\hat{q}_{{\cal M}}={\hat q}-{\hat q}_{{\cal F}} \ \ \ {\rm and} \ \ \ 
\hat{p}_{{\cal M}}={\hat p}+{\hat p}_{{\cal
F}}.
\end{equation}
We recognize in the relations (\ref{epr}) the position and the momentum
observables  of a composed system introduced and discussed  first by
Einstein, Podolsky and Rosen in their famous paper on the completeness of quantum
mechanics \cite{epr}. 

In our case the filter particles provide 
an example of the Naimark extension of ${\hat q}$ and ${\hat p}$. In the
extended space, the filter readout variables commute:
\begin{equation} 
[\hat{q}_{{\cal M}},\hat{p}_{{\cal M}}] = 
[{\hat q}-{\hat q}_{{\cal F}},{\hat p}+{\hat p}_{{\cal F}} ] = 0,
\end{equation}
and, as a result, can be measured simultaneously.
From this definition, we can introduce in this extended space 
a time observable:
\begin{equation} 
\hat{T}_{{\cal M}}= \frac{\hat{q}_{{\cal M}}}{\hat{p}_{{\cal M}}}\,,
\end{equation}
where the ordering of the commuting measured observables is irrelevant.
This expression will be central in the following investigations. 
The reduction of this operator to the Hilbert space ${\cal H}$ of the 
measured particle 
will provide an operational definition of the time of arrival 
in  phase space. As an example we obtain in the space of the particle
the following operator:
\begin{equation} 
\hat{T}= {\rm Tr}_{{\cal F}}
\left\{\frac{\hat{q}_{{\cal M}}}{\hat{p}_{{\cal M}}}\right\}.
\end{equation}

Using the definition of the propensity (\ref{propensity}) from  Section
II (with $a$ being $q$ and $p$), and the above interaction Hamiltonian we 
obtain that the propensity is a phase space distribution:
\begin{equation}
\Pr(q,p)=\frac{1}{2\pi}
\left|\int_{-\infty}^{\infty}\,dx \exp{(-ip x)}\psi(x){\cal
F}^{\ast}(q+x)\right|^2\,,
\end{equation}
where $\psi(x)$ and ${\cal F}(x)$ are the wave functions of the particle and
the filter  in the position representation labeled by $x$. In fact the filter
wave function is an overlap of the wave functions of the two particles $1$ 
and $2$ forming the measuring device.
The POVM in the Hilbert space of the system that corresponds to this 
measurement scheme is
\begin{equation}
\label{timePOVM}
\hat{F}(q,p)=\frac{1}{2\pi}
\exp{(iq\hat{p}-ip\hat{q})}|{\cal F}\rangle\langle
{\cal F}|
\exp{(-iq\hat{p}+ip\hat{q})}\,.
\end{equation}

\subsection{Mean time of arrival}

We shall apply this measurement to probe the time of arrival of a freely
moving particle probed by a filter device. In this case the mean time of
arrival will be simply related to the mean  relative position of the
measured particle with respect to a fixed filter position. In order to
describe such a measurement we shall apply the derived above phase space
distribution.

In such a case the propensity becomes time dependent via the
wave function $\psi(x,t)$ of the measured system. In order to carry all the
calculations
in an analytic form, we shall assume that  the time
measurement is performed on a freely moving particle described by the 
following Gaussian wave function:
\begin{equation}
\psi(x,t)=\left(\frac{\delta^2}{8 \pi}\right)^{1/4}
\frac{2\exp{(-\frac{1}{4}k_{0}^2\delta^2)}}{\sqrt{\delta^2+2it}}
\exp{
\left(\frac{(\frac{1}{2}k_{0}\delta^2+i(x-x_{0}))^2}{\delta^2+2it}
\right)},
\end{equation}
characterized by the initial mean position $x_{0}$, the initial mean 
momentum $k_{0}$
and width $\delta$.  The filter wave function, at the measurement time
$(t=0)$, is assumed to be in a stationary state with a Gaussian profile
centered around $q_{0}$ with width $\sigma$
\begin{equation}
{\cal F}(x)=\left(\frac{2}{\pi\sigma^2}\right)^{1/4}
e^{-(x-q_0)^{2}/\sigma^2}.
\end{equation}
As it is seen, due to the pulsed interaction between the filter and the
particle, the dynamics of the filter particle is unimportant.

For this choice of  the two wave packets the time evolution of the
 propensity is easily calculated
\begin{eqnarray}
\nonumber
\Pr(q,p,t) & = & \frac{1}{\pi}
\frac{\sqrt{\delta^2 \sigma^2}}{\sqrt{4t^2+(\delta^2+\sigma^2)^2}}
\exp{\left(-{
\frac{1}{2}\delta^2 \sigma^2 (\delta^2+ \sigma^2)(p-k_{0})^2
\over 4t^2+(\delta^2+\sigma^2)^2}\right)}  \\
&& \times \exp{\left(
{-2\left(\delta^2 (q-q_{0}+x_{0}+k_{0}t)^2+
\sigma^2 (q-q_{0}+x_{0}+pt)^2\right)
\over 4t^2+(\delta^2+\sigma^2)^2}
\right)}\,.
\end{eqnarray}
To understand the meaning of the propensity associated with
this measurement we calculate its first two phase space moments
\begin{mathletters}
\label{allequations}
\begin{equation}
\overline{q(t)}=q_{0}-(x_{0}+k_{0}t)\,,
\label{propm1}
\end{equation}
\begin{equation}
\overline{q(t)^2}=\overline{q}^{2}+\frac{t^2}{\delta^2}+
{\delta^2+\sigma^2 \over 4}\,,
\label{propm2}
\end{equation}
\begin{equation}
\overline{p}= k_{0}\,,
\label{propm3}
\end{equation}
\begin{equation}
\overline{p^2} = k_{0}^2+\frac{1}{\delta^2}+\frac{1}{\sigma^2}\,.
\label{propm4}
\end{equation}
\end{mathletters}
We see, for instance, that the average $\overline{q(t)}$ gives the
relative position of a particle measured with respect
to the filter position. 
This fact reflects the fundamental property of the measurement in which the
arrival time to a fixed position is monitored by a filter device.

\begin{figure}[t]
\begin{center}
\begin{tabular}{rcl}
\leavevmode  \epsfxsize=9.0cm \epsffile{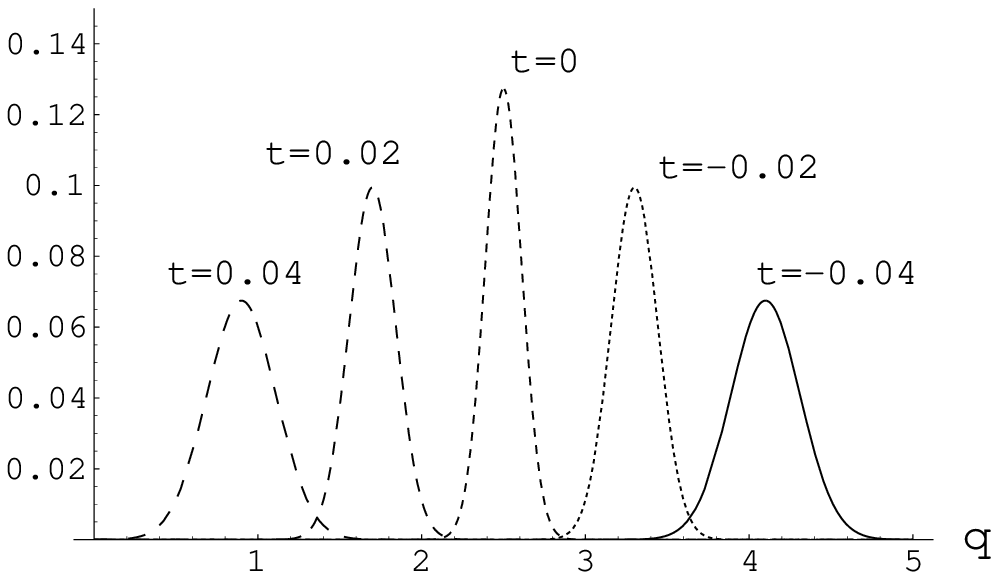} \\
\end{tabular}
\end{center}
\caption{Plot of the evolution of the cross section of the propensity 
$\Pr(q,p=40)$ for momentum $p=40$ at different times, with
$q_{0}=3.5$, $x_{0}=1$, $\delta=0.1$, $\sigma=0.1$, $k_{0}=40$ in
units $\hbar=m=1$.}
\label{propevol}
\end{figure}

The cross section of the propensity in $q$  for
a fixed value of $p$, as a function of $t$ is depicted on  Fig.~\ref{propevol}.
In this case  the particle moves from
a point $x_{0}$ towards the detection point $q_{0}$ of the filter.
Note a full symmetry between  positive and negative values of $t$, around
the measurement time set at $t=0$. With the passage of time, the distance
between the
measured particle and the filter particle is shrinking,
and finally it is zero, meaning that the measurement corresponding to the
arrival time at the point $q_{0}$ has been performed. 
Due to the unavoidable dispersions of the 
particle and the filter wave packets the propensity is spreading in time, 
deteriorating  the measurement precision.
The accuracy of such a measurement might be improved
in several ways. We can use well prepared 
wave packets of the system i.e., wave packets with a small dispersion, or 
we may prepare the filter state in a squeezed state \cite{sqstates},
with a reduced position dispersion. The measured momentum
is constant in time, which is a unique feature of the
Hamiltonian with a pulsed (Dirac delta function) interaction in time.

\section{Time Observable}

\subsection{Operational time of arrival}

From the phase space propensity derived in the previous section it is 
straightforward to define a mean arrival time and its moments using the
following definition:
\begin{equation}
\overline{\left(\frac{q}{p}\right)^n}=
\int_{\Gamma} dqdp\left(\frac{q}{p}\right)^n \Pr(q,p,t)\,,
\end{equation}
where the integration region $\Gamma$ is the whole phase space.
According to the general relation between the moments (\ref{moments}) and the
operational operators (\ref{op_operator}), we introduce a family of
operational time operators:
\begin{equation}
\langle\hat{T}^{(n)}\rangle\equiv \overline{\left(\frac{q}{p}\right)^n}
\end{equation}
where in terms of the operational time POVM (\ref{timePOVM}):
\begin{equation}
\label{timedef}
\hat{T}^{(n)} \equiv \int_{\Gamma} dqdp\left(\frac{q}{p}\right)^n
\hat{F}(q,p)\,.
\end{equation}
We notice immediately that for our choice of the
propensity the $p$-integral is divergent due to a singularity at $p=0$. This
singularity
can be easily understood on the physical ground, related to the concept of
the arrival time discussed above.
The physical reason for the singularity is that  
the particle with zero momentum will never
arrive to the measurement point, implying an infinite arrival time.

Obviously if we change the measurement  scheme in such a way that
(introducing for example a proper  external potential) 
the particle is forced to move towards the detection
point, the infinity in (\ref{timedef}) would be
removed.  In fact such a divergence is not unusual in quantum mechanics,
when for instance, vectors of the Hilbert space corresponding to an
infinite mean value of the position operator are used. For the time operator
operationally based  on the measurement of the arrival time, we shall select
only such states for which  the probability of finding the
particle with zero momentum  is very small.

In order to remove  the   
divergence we can apply a regularization procedure,
removing from the integration region the values of $p$ near zero.  As a result
we obtain:
\begin{equation}
\label{timedefreg}
\hat{T}^{(n)}_{\varepsilon}=
\int_{\Gamma_{\varepsilon}} dqdp\left(\frac{q}{p}\right)^n \hat{F}(q,p)\,,
\end{equation}
where for $\varepsilon > 0$
$\Gamma_{\varepsilon}=\{(q,p) \mid q\in {\Bbb R},
p\in{\Bbb R} \setminus [-\varepsilon,\varepsilon]\}$.
A similar regularization for a differently ordered time operator
was described in great detail in Ref. \cite{ng+cr+rst96}.
Our conclusion is analogous to the results of this paper. If the initial
wave packet is prepared in such a way that it has a 
vanishing distribution of momentum near $p=0$, 
the final results involving the time operator
are independent of the regularization procedure.
The condition which must be fulfilled is that
$|k_{0}|\delta\gg 1$. 

\begin{figure}[t]
\begin{center}
\begin{tabular}{rcl}
\leavevmode  \epsfxsize=9.0cm \epsffile{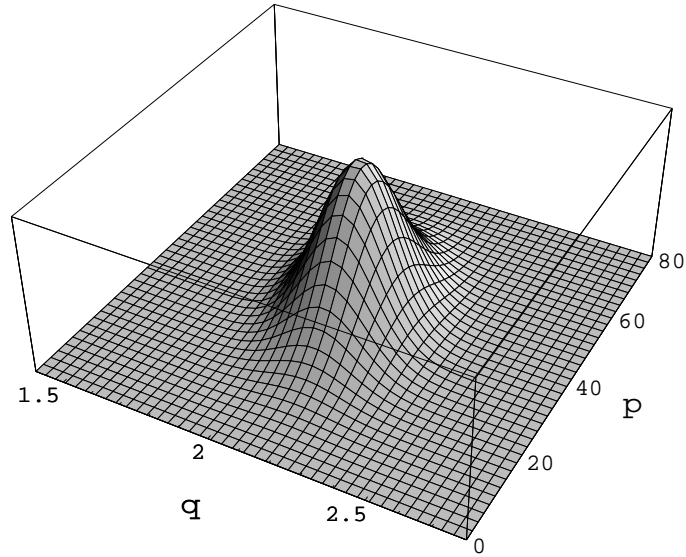} \\
\end{tabular}
\end{center}
\caption{Plot of the propensity $\Pr(q,p)$ at the time $t=0.01$,
with $q_{0}=3.5$, $x_{0}=1$, $\delta=0.1$, $\sigma=0.1$, $k_{0}=40$
in units $\hbar=m=1$.
Near  $p=0$, the propensity is  practically zero justifying
our approximation.}
\label{prop001}
\end{figure}

On Fig.~\ref{prop001} we have depicted the phase space propensity for a
selected set of values of $t,q_{0},x_{0},\delta,\sigma$ and $k_{0}$.
From this plot of the propensity, 
we see that it is enough to set, for example,
$k_{0}=40$ and $\delta=0.1$ (in our units) to
have a propensity practically equal to zero in
the required region. It is clear that if both the
particle wave function and the filter wave function are
badly localized, the notion of the time of arrival looses its
meaning. In this case a different time operator ought to be introduced.

Because the only significant contribution in the phase space 
integration comes from the
momenta concentrated around $p=k_0$ we can expand $1/p$ in
(\ref{timedefreg}) in the series around $k_{0}$,
(from a mathematical point of view such an expansion
of the integrand will lead usually, after integration, to
an asymptotic series). This greatly simplifies the
integration in (\ref{timedefreg}), and as a result we obtain  the following
leading contributions to the two first moments:
\begin{mathletters}
\label{talleq}
\begin{equation}
\langle\hat{T}^{(1)}\rangle=
\left(\frac{q_{0}-x_{0}}{k_{0}}-t\right)+
{\cal O}
(1/(k_{0}^2 \delta^2),1/(k_{0}^2 \sigma^2),1/(k_{0}^2 \delta \sigma))\,,
\label{t1}
\end{equation}
\begin{equation}
\langle\hat{T}^{(2)}\rangle=\langle\hat{T}^{(1)}\rangle^2 +
{\delta^2 + \sigma^2 \over 4k_{0}^2} +
{\cal O}
(1/(k_{0}^2 \delta^2),1/(k_{0}^2 \sigma^2),1/(k_{0}^2 \delta \sigma))\,,
\label{t2}
\end{equation}
\end{mathletters}
As we see, the leading contribution reproduces  the semiclassical
approximation of the  motion of the particle. The physical interpretation 
of the obtained results is clear. The mean of, the introduced time operator, 
gives the time of flight between the two points
$x_{0}+k_{0}t$ and $q_0$.  This time of arrival is given as a ratio of the
distance between two points and the particle momentum. Obviously there are
quantum corrections to the formulas (\ref{talleq}). These corrections are
much more pronounced when a second moment of the time
operator is investigated.  

\subsection{Ordering and time of arrival}

We shall derive now a more direct formula for the operational time
operator (\ref{timedef}) in the case of a particular choice of the
filter wave packet parameters. We express the position and the momentum
operators, in terms of the oscillator creation $\hat{a}^{\dagger}$ and
annihilation $\hat{a}$ operators, using the standard relations  
${\hat q}=(1/\sqrt{2})(\hat{a}^{\dagger}+\hat{a})$ and
${\hat p} =(i/\sqrt{2})(\hat{a}^{\dagger}-\hat{a})$. 
We set the filter wave function to be a  state  corresponding to 
$\langle q|\hat{a}|{\cal F}\rangle =0$. Such a state is a Gaussian wave
function with 
dispersion $\sigma=\sqrt{2}$ in our units). 
Using this filter state and these operators we
can rewrite the POVM as follows
\begin{equation}
\hat{F}(\alpha)=\frac{1}{\pi}\hat{D}(\alpha)
|0\rangle\langle 0|\hat{D}^{\dagger}(\alpha)
\equiv |\alpha\rangle\langle \alpha|
\ \ {\rm with} \ \ \int d^{2} \alpha 
\,\hat{{\cal F}}(\alpha)=\hat{1},
\end{equation}
where
$\hat{D}(\alpha)\equiv\exp{(\alpha\hat{a}^{\dagger}-\alpha^{\ast}\hat{a})}$
is the Glauber displacement  operator in the complex 
$\alpha=(1/\sqrt{2})(q+ip)$ phase space \cite{glauber63}. 
In such case, i.e., for such a filter wave function, the propensity  is
simply $P(\alpha) = \frac{1}{\pi} 
\langle\alpha | {\hat \rho} |\alpha \rangle$. We recognize
in this expression the    so called $Q$~--representation used  in quantum
optics applications.
It is well known that this function is related to the antinormal ordering of
the creation and annihilation operators 
($\hat{a}^{\dagger}$ to the right and
$\hat{a}$ to the left) \cite{glauber63}.

The time observables then take a very simple form 
\begin{equation}
\label{timeanty}
\hat{T}^{(n)}=\vdots \, \left(\frac{\hat{q}}{\hat{p}}\right)^n \,\vdots\,,
\end{equation}
where  the triple dots  denote the antinormal order of
the creation and  annihilation operators.

We see that the   measurement scheme  with
this particular choice of the filter wavepacket  leads
naturally to an antinormal ordering of
the creation and annihilation operators, which reflects 
a  specific ordering of the momentum and  position operators forming the
operational arrival time observables (\ref{timeanty}).
For a different choice of the width parameter $\sigma$, we will
have a different ordering of these operators (see \cite{ss92} for details).

As it was discussed above, only these states, which have
vanishing amplitudes for  momenta equal to zero are
physically interesting, and for such states the time operator
written above is well defined. Such states  specify the domain of the
operator (\ref{timeanty}).

In that spirit we will find a more explicit formula for the first
two moments of the time operator. We need to express the inverse
of antinormally ordered (in the discussed above sense)
momentum operator as a direct function of momentum operator centered 
around $k_{0}\neq 0$:
\begin{equation}
\vdots\frac{1}{\hat{p}}\vdots=\frac{1}{k_{0}}\vdots 
{1 \over 1+\frac{\hat{p}-k_{0}}{k_{0}}}\vdots=
\frac{1}{k_{0}}\sum_{n=0}^{\infty}(-1)^n 
\vdots\left({\hat{p}-k_{0}\over k_{0}}\right)^n\vdots\,.
\end{equation}
In this formula $k_{0}$ will be understood as mean value of momentum of 
a detected particle, and that states with zero momentum are excluded. 
Naturally, the above equality
holds only when we assure that there is a domain of vectors 
${\cal D}=\{\psi\in {\cal H}\mid ||(\hat{p}/k_{0}-1)\psi|| <1$\}. 
All the following formulas will
have a physical meaning for states belonging to the this domain.

In the next derivations we shall use the following algebraic identity:
\begin{equation}
\vdots \exp( \lambda \hat{p})\vdots = 
\exp\left( \frac{\lambda^{2}}{4} + \lambda
\hat{p}\right),
\end{equation}
where $\lambda$ is an arbitrary parameter. Using this formula simple algebra
gives
\begin{equation}
\vdots\frac{1}{\hat{p}}\vdots=
\frac{1}{k_{0}}\sum_{n=0}^{\infty}
\left({-1 \over 2ik_{0}}\right)^n {\rm H}_{n}(i(\hat{p}-k_{0}))\,,
\end{equation}
where ${\rm H}_{n}(z)$ are Hermite polynomials.  In order to find a second
moment of a time operator we need the following expression
\begin{equation}
\vdots\frac{1}{\hat{p}^2}\vdots=
\lim_{\epsilon\rightarrow 0} \frac{\partial}{\partial\epsilon}
\vdots\frac{1}{\hat{p}-\epsilon}\vdots=
\frac{1}{k^{2}_{0}}\sum_{n=0}^{\infty}
\left({-1 \over 2ik_{0}}\right)^n (n+1) {\rm H}_{n}(i(p-k_{0}))\,.
\end{equation}
All higher powers $\vdots (1/\hat{p})^n\vdots$ are easily obtained
by the subsequent derivation over $\epsilon$, and then taking the
limit $\epsilon\rightarrow 0$, as shown above for $n=2$.

Combining these algebraic properties we obtain the following formulas for the first two moments
of the operational time operator:
\begin{eqnarray}
\hat{T}^{(1)}&=&\frac{1}{2}\left(\frac{\hat{q}}{k_{0}}
\sum_{n=0}^{\infty}
\left({-1 \over 2ik_{0}}\right)^n {\rm H}_{n}(i(\hat{p}-k_{0}))
+\sum_{n=0}^{\infty}
\left({-1 \over 2ik_{0}}\right)^n {\rm H}_{n}(i(\hat{p}-k_{0}))
\frac{\hat{q}}{k_{0}}\right)\,, \\
\hat{T}^{(2)}&=&\frac{1}{2}\left(
\vdots\frac{1}{\hat{p}^2}\vdots+
\frac{1}{2}\hat{q}^2\vdots\frac{1}{\hat{p}^2}\vdots+
\frac{1}{2}\vdots\frac{1}{\hat{p}^2}\vdots \hat{q}^2+
\hat{q}\vdots\frac{1}{\hat{p}^2}\vdots\hat{q}\right)\,,
\end{eqnarray}
where $\vdots 1/\hat{p}^2\vdots$ was calculated above.

In the subspace ${\cal D}$ an arbitrary moment of the operational time
operator can be written in the following form:
\begin{equation}
\hat{T}^{(n)} =\sum_{k=0}^{\infty} \hat{T}_{[k]}^{(n)}
\end{equation}
where the subscript index is of order 
$( ({\hat p}-k_{0})/k_{0}) )^{k+1}$, and a
systematic method of calculating  all $\hat{T}_{[k]}^{(n)}$ follows from
the algebra presented above.
For sufficiently large values of $k_{0}$ i.e., for a fast
moving particle we can well approximate the formula for the
time operator by the  first few terms in the series.
For $\hat{T}^{(1)}$ we have:
\begin{equation}
\hat{T}^{(1)} = \hat{T}_{[0]}^{(1)} + \hat{T}_{[1]}^{(1)} +
\hat{T}_{[2]}^{(1)} +{\cal O}\left(\left(\frac{\hat{p}-k_{0}}{k_{0}}\right)^3,
\frac{1}{k^{3}_{0}}\right)\,, 
\end{equation}
where
\begin{mathletters}
\label{allequations1}
\begin{equation}
\hat{T}_{[0]}^{(1)}={\hat{q} \over k_{0}}\,,
\label{t10}
\end{equation}
\begin{equation}
\hat{T}_{[1]}^{(1)}= -
\frac{1}{2}\left({\hat{q} \over k_{0}}\frac{\hat{p}-k_{0}}{k_{0}}+
\frac{\hat{p}-k_{0}}{k_{0}}{\hat{q} \over k_{0}}\right)\,,
\label{t11}
\end{equation}
\begin{equation}
\hat{T}_{[2]}^{(1)}= \frac{1}{2}\left({\hat{q} \over
k_{0}}\left(\frac{\hat{p}-k_{0}}{k_{0}}\right)^2+
\left(\frac{\hat{p}-k_{0}}{k_{0}}\right)^2 {\hat{q} \over k_{0}}\right)\,.
\label{t12}
\end{equation}
\end{mathletters}
The zero order term reproduces the classical limit corresponding to an
arrival time of a particle with momentum $k_{0}$ at a fixed position $q$.

For $\hat{T}^{(2)}$ we have

\begin{mathletters}
\label{allequations2}
\begin{equation}
\hat{T}^{(2)}_{[0]}=\frac{\hat{q}^2}{k_{0}^{2}}+
\frac{1}{2k_{0}^2},
\label{t20}
\end{equation}
\begin{equation}
\hat{T}^{(2)}_{[1]}=
-\frac{1}{k_{0}^2}\left( \frac{\hat{p}-k_{0}}{k_{0}}+ 
\hat{q}\frac{\hat{p}-k_{0}}{k_{0}}\hat{q}\right) 
-\frac{1}{2k_{0}^2}\left( 
\hat{q}^2\frac{\hat{p}-k_{0}}{k_{0}}+\frac{\hat{p}-k_{0}}{k_{0}}\hat{q}^2
\right)\,.
\label{t21}
\end{equation}
\end{mathletters}
Even in the zero order approximation the second moment of the
time observable {\em does not} reproduce the classical limit, in
particular we see that $(\hat{T}^{(1)})^2\neq\hat{T}^{(2)}$.

Usually when we deal only with mean values of observables, the difference
between results obtained with the help of an intrinsic operator
(if such exists) and the operational operator is rather simple and
straightforward. The first
moment of the operational operator equals to the intrinsic operator, or
differs by a term simply connected to the properties of the  of the
filter device.
The situation becomes quite different for
the second moments. In the operational formalism the second moment contains
a contribution representing quantum  fluctuations of the filter. Due to the 
additive character
of the presented measurement the time operator $\hat{T}^{(2)}$ contains 
an additive quantum noise contribution from the filter system.

We conclude this section investigating the eigenfunctions of the 
time operator.
The eigenvalue problem for the operational time of arrival has the following
form: \begin{equation}
\hat{T}^{(1)}|\chi_{\tau}\rangle = \tau |\chi_{\tau}\rangle.
\end{equation}
In the following we derive the eigenvectors for several approximated values
of $\hat{T}^{(1)}$. In the zero approximation, for $\hat{T}_{[0]}^{(1)}$, the
eigenstate can be  simply calculated in the momentum representation. For a
given 
eigenvalue $\tau$ this eigenvector is just
\begin{equation}
\chi^{(0)}_{\tau}(p)= \langle p|\tau\rangle =e^{-i\tau k_{0}(p-k_{0})}\equiv
e^{-iT k_{0}^2}\,,
\end{equation}
where we have introduced a new variable $T=\tau(p-k_{0})/k_{0}$.
As it is seen, the eigenstates are labelled by the momentum
of the inspected particle. If we treat $\tau$ as a physical time,
we can write $\tau p\equiv q$, $\tau k_{0}\equiv q_{0}$ and
then $T$ can be interpreted as the time of flight from the point
$q_{0}$ to a fixed point $q$. The spectrum of this operator is continuous.

The eigenfunctions of the time operator involving the higher order
contribution $\hat{T}_{[1]}^{(1)}$ can be obtained  solving a  first order
differential equation. As a result we obtain:
\begin{equation}
\chi^{(1)}_{\tau}(p)=
{\exp{\left(iTk_{0}\log{\left(1-\frac{p-k_{0}}{k_{0}}\right)}\right)}
\over
\sqrt{1-\frac{p-k_{0}}{k_{0}}}}\,.
\end{equation}
This eigenfunction will belong to the domain ${\cal D}$, if $(p-k_{0})/k_{0}
< 1 $. The connection with the
eigenfunction in the lower order of approximation and
the interpretation of this result is best seen upon expanding
the eigenfunction in powers of the parameter
$(p-k_{0})/k_{0}$, we have then
\begin{equation}
\chi^{(1)}_{\tau}(p)\approx
\left(1+\frac{1}{2}\frac{p-k_{0}}{k_{0}}+\ldots\right)
\exp{\left(-ik^2_{0}T\left(1+\frac{1}{2}\frac{p-k_{0}}{k_{0}}+
\frac{1}{3}\left(\frac{p-k_{0}}{k_{0}}\right)^2+\ldots\right)\right)}\,,
\end{equation}
where, as above, $T=\tau(p-k_{0})/k_{0}$. 
From this expression we conclude that the phase of
this eigenfunction is the classical arrival time $T$ with  nonclassical
contributions resulting from the higher order term $\hat{T}_{[1]}^{(1)}$.

\section{Operational time-energy uncertainty}

In the approach presented in this paper it is  easy to give a clear 
meaning to the
time and  energy uncertainty relation. This is because that, on the 
operational level, we can describe a measurement of both the time of arrival
for a given system and its energy.
The operational time operator is given by Eq. (\ref{timedef}), and the
corresponding operational moments of energy of the system are:
\begin{equation}
\label{energy}
\hat{E}^{(n)} \equiv \int dp \ \left(\frac{p^{2}}{2}\right)^{n}
\hat{F}(p)\,,
\end{equation}
where $\hat{F}(p) = \int dq \hat{F}(q,p)$ is the momentum marginal of the
phase space POVM (\ref{timePOVM}).
It should  be pointed out that the defined below uncertainty relation
is given between the {\em measured} time of arrival
and the {\em measured} energy of the chosen system, and as
a result, the criticism of Aharonov and Bohm \cite{ya+db6164} and
Peres \cite{peres} concerning those
approaches where one tried to construct the uncertainty principle
between the time of arrival detected by the filter and the energy
of the investigated particle (two commuting quantities)
does not apply here.

Using the general formula for the operational spread introduced in Section II,
we introduce the operational spread of the time operator 
\begin{equation}
\delta T \equiv \langle \hat{T}^{(2)} \rangle -
\langle \hat{T}^{(1)} \rangle^{2}.
\end{equation}
This quantity can be  calculated using Eq.(\ref{t2}). 
The corresponding operational spread of the energy is:
\begin{equation}
\delta E \equiv \langle \hat{E}^{(2)} \rangle -
\langle \hat{E}^{(1)} \rangle^{2}. 
\end{equation}
The mean operational energy $\langle \hat{E}^{(1)} \rangle$ 
has been already calculated
(\ref{propm4}) and we need only
\begin{equation}
\langle \hat{E}^{(2)} \rangle =\frac{1}{4}\left[ k_{0}^4+
6 k_{0}^2 \left( {\delta^2 + \sigma^2 \over \delta^2 \sigma^2}\right)+
3 \left({\delta^2 + \sigma^2 \over \delta^2 \sigma^2}\right)^2\right]\,.
\end{equation}
Simple algebra gives,as a result, an operational time-energy uncertainty in the form:
\begin{equation}
\delta E\ \delta T \equiv
(\langle E^{(2)}\rangle-
(\langle E^{(1)} \rangle^2)
(\langle\hat{T}^{(2)}\rangle-\langle\hat{T}^{(1)}\rangle^2
\,)
\geq
\frac{1}{2}{\delta^2 + \sigma^2 \over \delta \sigma}\,.
\end{equation}
When both $\delta$ and $\sigma$ are approaching simultaneously
zero, the right
hand side of the above inequality tends to $1$ (i.e., $\hbar$), meaning
that the ``conventional'' uncertainty relation is reproduced
if the particle and filter states are very well prepared.
However, we have to remember that these calculations have been performed 
under the assumptions that $k_{0}\delta \gg 1$ and 
$k_{0}\sigma \gg 1$. This corresponds to the position spread of  the
wavepackets  very small and hence giving the momentum spread very large.

We conclude this section calculating the commutator involving the
energy ${\hat E}^{(1)}$ and the operational time operator ${\hat T}^{(1)}$ 
derived for a filter state leading to an
antinormal form. As it was said
before,
these first moments of the operational operators should in some sense be the
closest to the intrinsic time and energy observables. A simple algebra shows
that:
\begin{equation}
[{\hat T}^{(1)},{\hat E}^{(1)}]= i  \left( 1 + 
\frac{{\hat p} -k_{0}}{k_{0}}\right) \vdots 
\frac{1}{1 + \frac{{\hat p} -k_{0}}{k_{0}} }\vdots \,. 
\end{equation}
The right hand side of this commutator is complicated, but for states
belonging to the domain ${\cal D}$ we obtain that the leading term is:
\begin{equation}
[{\hat T}^{(1)},{\hat E}^{(1)}]= i  + {\cal O} \left( \frac{{\hat p}
-k_{0}}{k_{0}}\right).
\end{equation}
Only in this limit, one can associate the traditional interpretation of
${\hat T}^{(1)}$ and ${\hat E}^{(1)}$ as canonically conjugated variables 
for time and energy.

\section{Time and Energy Phase Space}

\subsection{Probability distribution of time of arrival}

The definition of the operational time of arrival
operator is related to the following moments $\overline{(q/p)^n}$ of
the corresponding phase space propensity. 
It seems natural then to introduce
a differently parameterized propensity i.e.,  instead of working
with $\Pr(q,p,t)$ as a function of $q$, $p$ we shall define a new
 arrival ``time'' variable $\theta=q/p$ (to be not confused with
the time $t$ labelling the evolution of the system wave function):
\begin{equation}
\Pr(\theta,t)=\overline{\delta\!\!\left(\frac{q}{p}-\theta\right)}
\equiv \int\, dq dp\, \delta\!\!\left(\frac{q}{p}-\theta\right) \Pr(q,p,t)\,.
\end{equation}
From the properties of the Dirac delta function we obtain
\begin{equation}
\label{op_flux}
\Pr(\theta,t)=
\int_{0}^{\infty} dp\, p\Pr(\theta p,p,t) +
\int_{0}^{\infty} dp\, p\Pr(-\theta p,-p,t) \equiv 
{\Pr}_{+}(\theta,t)+ {\Pr}_{-}(\theta,t)\,.
\end{equation}
The interpretation of this result is simple in terms of the right $(+)$
and left $(-)$ operational probability flux.  

This result should be contrasted with the standard quantum  
current operator      
\begin{equation}
\label{qmflux}
\hat{j}(x) = \frac{1}{2}\left( \hat{p} |x\rangle\langle x| +
|x\rangle\langle x|\hat{p}\right)\,.
\end{equation}
The expectation value of this current seems to be
a natural and intuitive candidate for the probability distribution of 
the time
of arrival at the given space point $x$. 
It is however difficult to make a practical use of this intuition
since the flux (\ref{qmflux}) is nonpositive and it is difficult 
to associate with such a nonpositive flux a probability distribution.
A  discussion of this problem has been  
presented by Delgado \cite{vd97+98}. 

Because 
the position and the momentum of the particle cannot be specified with
arbitrary precision there is a possibility for a quantum backflow 
contribution to  $\langle \hat{j}(x) \rangle $, if it is understood as a
probability distribution. As it was pointed out in Ref. \cite{vd97+98},
when this backflow part of the probability distribution becomes
negligible one can use this formula to define the time observable.
Indeed, if a wavepacket is constructed in such way
that the mean value of the momentum is positive and large,
and a small negative part of the flux can be neglected,
the probability distribution for an arrival time is approximately
positive.
However, this backflow contribution is interesting in itself
\cite{jgm+jpp+crl} since it is of purely quantum mechanical origin. This
is why it is tempting not to exclude it from the flux considerations, and 
in our approach based on the operational propensity, there is no need to do so.

\begin{figure}[t]
\begin{center}
\begin{tabular}{rcl}
\leavevmode  \epsfxsize=9.0cm \epsffile{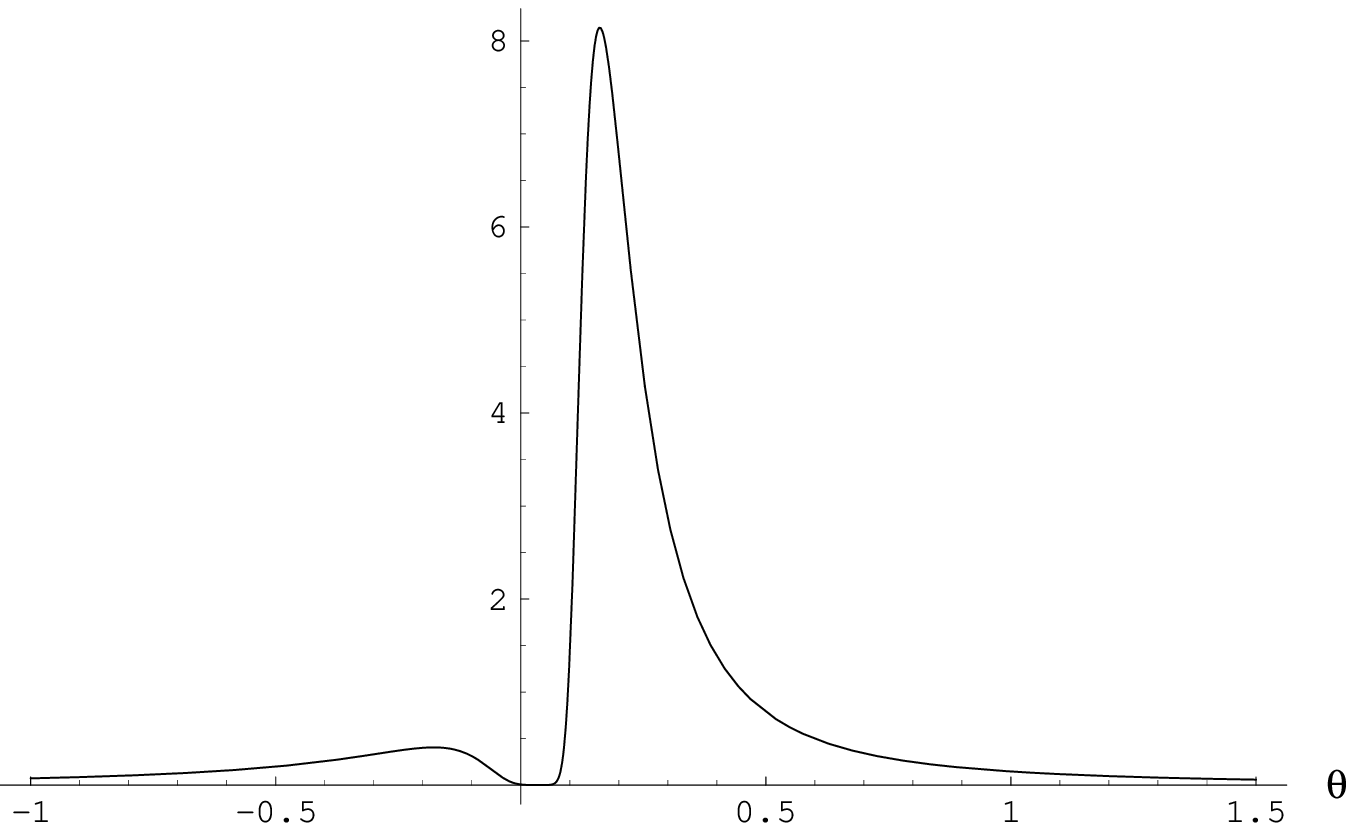} \\
\end{tabular}
\end{center}
\caption{Plot of the propensity $\Pr(\theta,t)$ as a function
of $\theta$ at $t=-0.1$, with
$q_{0}=3.5$, $x_{0}=1$, $\delta=0.1$, $\sigma=0.1$, $k_{0}=10$
in units $\hbar=m=1$.
The right hand part of the plot shows the ``positive'' momenta 
contribution, whereas on the left hand side part of the plot we see 
the contribution from the ``negative'' momenta ---the  backflow effect. }
\label{prt}
\end{figure}

Contrary to the quantum mechanical flux (\ref{qmflux}), the operational flux
given by (\ref{op_flux}) is clearly positive. And as we have seen
in the previous sections an association between this
flux and the time observable  is very natural and simple.  In fact
the flux (\ref{op_flux}) is exactly the corresponding  time of
arrival probability distribution,
\begin{equation}
\langle T^{(1)} \rangle =
\int d\theta \, \theta \Pr(\theta,t).
\end{equation}
Obviously, in our approach we deal with the two parts of the flux.
The ``positive'' part $\Pr_{+}(\theta,t)$ corresponds to the
particle moving in the direction given by the mean value
of the momentum $k_{0}$. The ``negative'' (or backflow) part has obviously
the opposite meaning. The presence of these two parts in
the flux is clearly a purely quantum feature, because in the classical
world the particle moves either to one or to the other direction
with probability one. For our choice of the wavepackets
the negative and the positive parts of the 
propensity (\ref{op_flux}) might be written
explicitly:
\begin{eqnarray}
\nonumber
{\Pr}_{\pm}(\theta,t)&=&\frac{1}{2\pi}
{\sqrt{\sigma^2 \delta^2} \sqrt{4t^2+(\sigma^2+\delta^2)^2}
\over \Delta(\theta,t)}
\left(2+\frac{\sqrt{\pi}\xi}{2} e^{\xi^2}({\rm erf}(\xi) \pm 1)
\right) \\
&&\times \exp{\left\{
{-2k_{0}^2 \left(\delta^2 T_{cl}(t)+\sigma^2 T_{cl}(0)+\Delta(0,0)\right)
\over 4t^2+(\sigma^2+\delta^2)^2}
\right\}}\,,
\end{eqnarray}
where
\begin{eqnarray}
T_{cl}(t)&=&\frac{q_{0}-(x_{0}+k_{0}t)}{k_{0}}\,,\,\,\,\,
T_{cl}(0)=\frac{q_{0}-x_{0}}{k_{0}}\,, \\
\Delta(\theta,t)&=&\sigma^2(\theta+t)^2+\theta^2\delta^2+
\frac{1}{4}\delta^2 \sigma^2 (\delta^2+\sigma^2),\,\,\,\,
\Delta(0,0)= \frac{1}{4}\delta^2 \sigma^2 (\delta^2+\sigma^2)\, \\
\xi&=&{2 k_{0} (\sigma^2(\theta+t)T_{cl}(0)+\delta^2\theta T_{cl}(t)+
\Delta(0,0)) \over
\sqrt{2\Delta(\theta,t)}\sqrt{4t^2+(\sigma^2+\delta^2)^2}}\,,
\end{eqnarray}
and  ${\rm erf(z)}$ is the error function.
When the mean momentum $k_{0}$ is large and positive the contribution of  
the negative
flux  $\Pr_{-}(\theta,t)$
becomes small. In this case the ratio of these two parts might me
approximated as follows:
\begin{equation}
\frac{\Pr_{-}(\theta,t)}{\Pr_{+}(\theta,t)}
\approx \frac{3}{2\sqrt{\pi}}\left(\frac{1}{\xi}+
{\cal O}\left(\frac{1}{\xi^2}\right)\right)e^{-\xi^2}
\end{equation}
When the particle is moving fast, $k_{0}\gg0$, then $\xi$ becomes
large and the above ratio tends to zero.
Finally, the formula for the complete flux is given by
\begin{eqnarray}
\nonumber
\Pr(\theta,t)&=&\frac{1}{2\pi}
{\sqrt{\sigma^2 \delta^2} \sqrt{4t^2+(\sigma^2+\delta^2)^2}
\over \Delta(\theta,t)}
\exp{\left\{
{-2k_{0}^2 \left(\delta^2 T_{cl}(t)+\sigma^2 T_{cl}(0)+\Delta(0,0)\right)
\over 4t^2+(\sigma^2+\delta^2)^2}
\right\}}+ \\
&&\frac{1}{2\pi}
{\sqrt{\sigma^2 \delta^2} \sqrt{4t^2+(\sigma^2+\delta^2)^2}
\over \Delta(\theta,t)}
\xi{\rm erf(\xi)}
\exp{ \left\{ -\delta^2 \sigma^2(\theta-T_{cl}(t))^2 \over 2\Delta(\theta,t)
\right\}}\,.
\end{eqnarray}
This quantity is depicted  on Fig.~\ref{prt} for a particular choice
of the wave function parameters. As it's seen the backflow part
of the flux is much smaller then the dominating ``positive'' part.
What is important is the fact that in our considerations we do not have to
neglect the ``negative'' momenta contribution. In fact our model
measurement scheme allows to measure this part of the probability
distribution.

It is also interesting to note, that having defined the time observable
Delgado and Muga \cite{vd+jgm97,vd97+98}, also have managed to associate
with it a positively defined current. This  shows that such
a relation is quite universal.

In analogy  to the time distribution, we can define a complementary quantity, a probability
distribution connected with the energy measurement:
\begin{equation}
\label{energy_def}
\Pr(E)=\overline{\delta\left(\frac{p^2}{2}-E\right)}
\equiv \int\, dq dp\, \delta\left(\frac{p^2}{2}-E \right) \Pr(q,p,t)\,.
\end{equation}
Due to the pulsed interaction of the filter with the system, this
distribution is time independent. Indeed, a simple calculation shows that:
\begin{equation}
\Pr(E)=\frac{1}{\sqrt{\pi E}}
\sqrt{\frac{\delta^2 \sigma^2}{\delta^2+\sigma^2}}
\exp{\left\{- { \delta^2 \sigma^2 (k_{0}^2+E) \over
2(\delta^2+\sigma^2)}\right\}}
\cosh\left({ \delta^2 \sigma^2 (k_{0}\sqrt{2E}) \over
(\delta^2+\sigma^2)}\right)
\end{equation}

\subsection{Time and energy phase space}

\begin{figure}[t]
\begin{center}
\begin{tabular}{rcl}
\leavevmode  \epsfxsize=9.0cm \epsffile{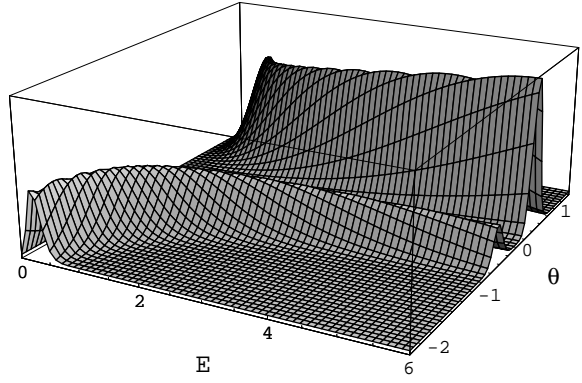} \\
\end{tabular}
\end{center}
\caption{Plot of joint propensity $\Pr(E,\theta)$
at the time $t=0.01$, with $q_{0}=3.5$, $x_{0}=1$, $\delta=0.1$, 
$\sigma=0.1$, $k_{0}=40$ in units $\hbar=m=1$. Again we see that there are two parts,
since there are contributions from both ``negative'' and ``positive''
momenta.}
\label{pret1}
\end{figure}

\begin{figure}[t]
\begin{center}
\begin{tabular}{rcl}
\leavevmode  \epsfxsize=9.0cm \epsffile{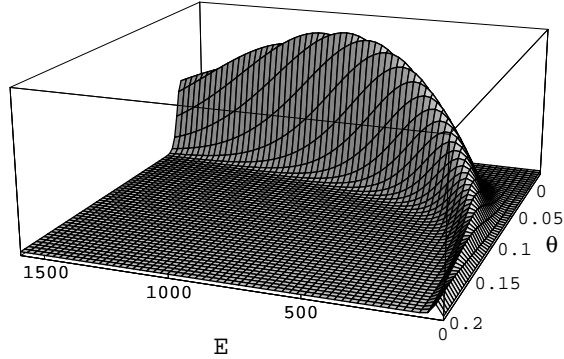} \\
\end{tabular}
\end{center}
\caption{Plot of joint propensity $\Pr(E,\theta)$
at the time $t=0.01$, with $q_{0}=3.5$, $x_{0}=1$, $\delta=0.1$, 
$\sigma=0.1$, $k_{0}=40$ in units $\hbar=m=1$. The range of that plot is 
chosen such that
we see only the positive momenta contribution. For low energies 
the time of arrival is very inaccurate --- the
propensity is very broad, but the measurement of energy is precise;
for large momenta the situation is reversed, which is the consequence
of the uncertainty relation for time and energy.}
\label{pret2}
\end{figure}

In conclusion of this section we shall introduce a combined time and energy
phase space for the operational measurement. In such a phase space we have a
joint energy and momentum distribution defined as:
\begin{equation}
\Pr(\theta,E)=\overline{\delta\left(\frac{p^2}{2}-E\right)
\delta\left(\frac{q}{p}-\theta\right)}\,.
\end{equation}
The  time distribution (\ref{op_flux}) and the energy
distribution (\ref{energy_def}) are marginals of this joint time and energy
distribution. 
For the Gaussian wave functions we calculate
\begin{eqnarray}
\Pr(\theta,E)&=&\frac{2}{\pi}
\nonumber
{\sqrt{\sigma^2 \delta^2} \over \sqrt{4t^2+(\sigma^2+\delta^2)^2}}
\cosh\left( { 4k_{0}\sqrt{2E}(
\delta^2 T_{cl}(t)\theta + \sigma^2 T_{cl}(0)(\theta+t)+\Delta(0,0))
\over 4t^2+(\sigma^2+\delta^2)^2}\right) \\
&&\times
\exp{\left\{ {- 2k^2_{0} \over 4t^2+(\sigma^2+\delta^2)^2}
\left[
\frac{2E}{k^2_{0}}\Delta(\theta,t)+
\delta^2 T_{cl}(t)+\sigma^2 T_{cl}(0)+\Delta(0,0)\right]\right\}}\,.
\end{eqnarray}
The meaning of this quantity becomes clear, when we look on
its plot depicted on Fig.~\ref{pret1}. We see the characteristic separation
into two parts, one corresponding to the positive momenta
(relative to  $k_{0}$) and a much smaller contribution
from the negative part. Further, we notice 
that for small values of energy the propensity is narrow in
the $E$ direction but rather broad in the $\theta$
direction. This shows that the time measurement looses its meaning 
for very slow particles (small $k_{0}$).
The situation changes in a complementary way when the energy is increased. 
This is depicted on  Fig.~\ref{pret2}, where the propensity
becomes very sharp in the $\theta$ direction. This  
assures a sharp  measurement of the time of arrival.
In this case the energy measurement becomes less accurate.

The fact, that we have these two limits of the propensity 
is of course  a manifestation
of the time-energy uncertainty principle.  

\section{Conclusions}

With the help of the operational formulation we have introduced an 
operational time observable associated with a specific measurement scheme.
Such an operational approach is not universal, in a sense,
that the specific form of the time observable depends on the quantum state 
of the filter device.  This operational approach 
allows for a natural and clear 
definition of the time of arrival. 
Our result is an intuitive quantum counterpart of the classical
time of flight  measurement   between two points.
Based on such an operational approach the time -- energy  uncertainty 
principle has been introduced.
The idealized 
measurement scheme allows to provide  a link between  the quantum 
mechanical flux of the operational propensity and the time observable. 
This has been done without neglecting the backflow part, which is very 
interesting. 

The time operator discussed in this paper is related to a specific measuring
scheme. The question remains what have we learned about the intrinsic time
operator from our discussion. Our view is that various fundamental time 
operators will have many features of the operator $T^{(1)}$. In fact we have
shown that in terms of the commutation relations, classical limit and 
the physical interpretation, this operator has many features of an intrinsic
time observable. However the problem is that we cannot provide a first 
principle derivation of this observable from some fundamental assumptions. 
The operational observable has been build using a specific detection 
procedure, but its overall properties should describe a intrinsic time 
operator in a reasonable way.

\section*{Acknowledgments}
It is pleasure for us to acknowledge numerous discussions
with Professor Jerzy Kijowski.
This work has been partially supported by the Polish KBN grant
No. 2 PO3B 118 12.

\end{document}